\documentclass[5p,twocolumn,times,number]{elsarticle}

\usepackage{graphicx}
\usepackage{amsmath}
\usepackage[amssymb]{SIunits}
\usepackage[colorlinks,hyperindex]{hyperref}

\usepackage{lineno}

\begin{document}

\begin{frontmatter}

\title{Mu2e calorimeter readout system}
\date{November, 27, 2018}

\author[a]{N.~Atanov}
\author[a]{V.~Baranov}
\author[f]{L.~Baldini}
\author[a]{J.~Budagov}
\author[f,e]{D.~Caiulo\corref{cor}}
\ead{davide.caiulo@pi.infn.it}
\author[f]{F.~Cei}
\author[e]{F.~Cervelli}
\author[b]{F.~Colao}
\author[b]{M.~Cordelli}
\author[b]{G.~Corradi}
\author[a]{Yu.I.~Davydov}
\author[f]{F.~D'Errico}
\author[e]{S.~Di Falco}
\author[b,i]{E.~Diociaiuti}
\author[e,f]{S.~Donati}
\author[b,j]{R.~Donghia}
\author[c]{B.~Echenard}
\author[f]{S.~Faetti}
\author[b]{S.~Giovannella}
\author[f]{S.~Giudici}
\author[a]{V.~Glagolev}
\author[h]{F.~Grancagnolo}
\author[b]{F.~Happacher}
\author[c]{D.G.~Hitlin}
\author[f]{L.~Lazzeri}
\author[b,d]{M.~Martini}
\author[b]{S.~Miscetti}
\author[c]{T.~Miyashita}
\author[e]{L.~Morescalchi}
\author[g]{P.~Murat}
\author[f]{D.~Nicol\`o}
\author[e]{E.~Pedreschi}
\author[k]{G.~Pezzullo}
\author[f]{G.~Polacco}
\author[c]{F.~Porter}
\author[e]{F.~Raffaelli}
\author[b,d]{M.~Ricci}
\author[b]{A.~Saputi}
\author[b]{I.~Sarra}
\author[f]{M.~Sozzi}
\author[e]{F.~Spinella}
\author[h]{G.~Tassielli}
\author[a]{V.~Tereshchenko}
\author[a]{Z.~Usubov}
\author[a]{I.I.~Vasilyev}
\author[c]{R.Y.~Zhu}

\cortext[cor]{Corresponding author}

\address[a]{Joint Institute for Nuclear Research, Dubna, Russia}
\address[f]{Dipartimento di Fisica dell'Universit\`a di Pisa, Pisa, Italy}
\address[e]{INFN Sezione di Pisa, Pisa, Italy}

\address[b]{Laboratori Nazionali di Frascati dell'INFN, Frascati, Italy}
\address[c]{California Institute of Technology, Pasadena, United States}
\address[d]{Universit\`a ``Guglielmo Marconi'', Roma, Italy}
 
\address[g]{Fermi National Laboratory, Batavia, Illinois, USA}
\address[h]{INFN Sezione di Lecce, Lecce, Italy}
\address[i]{Dipartimento di Fisica dell'Universit\`a di Roma Tor Vergata, Rome, Italy}
\address[j]{Dipartimento di Fisica dell'Universit\`a degli Studi Roma Tre, Rome, Italy}
\address[k]{Yale university, New Haven, USA}

\begin{abstract}
The Mu2e electromagnetic calorimeter is made of two disks of un-doped parallelepiped CsI crystals readout by SiPM. There are 674 crystals in one disk and each crystal is readout by an array of two SiPM. The readout electronics is composed of two types of modules: 1) the front-end module hosts the shaping amplifier and the high voltage linear regulator; since one front-end module is interfaced to one SiPM, a total of 2696 modules are needed for the entire calorimeter; 2) a waveform digitizer provides a further level of amplification and digitizes the SiPM signal at the sampling frequency of $200\ \mega \hertz$ with 12-bits ADC resolution; since one board digitizes the data received from 20 SiPMs, a total of 136 boards are needed.
The readout system operational conditions are hostile: ionization dose of $20\ \text{krads}$, neutron flux of $10^{12}\ \text{n}(1\ \text{MeVeq})/\centi\meter^2$, magnetic field of $1\ \text{T}$ and in vacuum level of $10^{-4}\ \text{Torr}$. A description of the readout system and qualification tests is reported.

\end{abstract}

\begin{keyword}
Mu2e Calorimeter \sep digitizer \sep front-end electronics \sep radiation tolerance

\PACS 29.40.Cs \sep 29.40.Gx 
\end{keyword}

\end{frontmatter}

\section{Introduction}
The Mu2e experiment at Fermilab searches for the charged-lepton flavor violating (CLFV) conversion of a muon into an electron in the field of an aluminum nucleus, with a distinctive signature of a mono-energetic electron of $104.967\ \mega \text{eV}$. The Mu2e goal is to improve by four order of magnitude the search sensitivity with respect to previous experiments.

The Mu2e detector \cite{Mu2eTDR} is composed of a straw-tube tracker, an electromagnetic calorimeter and an external veto for cosmic rays surrounding the solenoid. 

The calorimeter \cite{Calo} is an high granularity crystal calorimeter consisting of 1348 undoped $\text{CsI}$ crystals, arranged in two disks. Each crystal is coupled to two large-area UV-extended SiPMs arrays. Each array consists of two series of three SiPMs. The two series are connected in parallel by the front-end electronics to have a $\times 2$ redundancy. The calorimeter is in a $10^{-4}\ \text{Torr}$ vacuum and in a uniform magnetic field of $1\ \text{T}$.

The Mu2e calorimeter plays an important role in providing particle identification, a seed for track reconstruction and a fast online trigger filter. 

The calorimeter requirements to detect $105\ \text{MeV}$ electrons are a time resolution better than $0.5\ \nano \second$, an energy resolution $< 10\%$ and a position resolution of $1\ \centi\meter$.

Since the calorimeter will be exposed to a very intense flux, a high sampling rate is needed to digitize the fast SiPMs signal with an appropriate number of samples to have a good resolution of the pileup of incoming particles. This implies the signals have to be sampled with a minimum sampling frequency of $200\ \mega\hertz$ with at least 12-bits of resolution.

All the DAQ system is located inside the cryostat to limit the number of pass-through connectors. This solution complicates the design in terms of available space and accessibility in case of failure and power dissipation. The space limitation implies the choice of 20 ADC channels/board and the limitation about the access requires an highly reliable design.

In addition, the digitizers operate in a $1\ \tesla$ axial magnetic field and in an harsh radiation environment. GEANT4/MARS simulation shows that the digitizer components will absorb a total ionizing dose of $1.6\ \text{krad/year}$ and will suffer a displacement damage equivalent to the one produced by a flux of $0.5\ \times 10^{11}\ \text{n}(1\ \text{MeVeq})/\centi\meter^2$ \citep{Pronskikh}.

The simultaneous presence of radiation and high magnetic field imposes stringent constraints on the component choice.

\section{The readout electronics}
The front-end electronics (FEE) consists of two discrete and independent chips (Amp-HV), for each crystal, directly connected to the back of the SiPM pins. These provide the amplification and shaping stage, a local linear regulation of the bias voltage, monitoring of current and temperature on the sensors and a test pulse. Groups of 20 Amp-HV chips are controlled by a dedicated mezzanine board (MB) equipped with an ARM (Advanced RISC Machine) to distribute the LV and the HV reference values and set and read back the locally regulated voltages. Groups of 20 signals are sent differentially to a digitizer module.

The digitizer module (DIRAC, DIgitizer and RedAdout Controller) provides a further level of amplification and digitizes the SiPM signal at the sampling frequency of $200\ \mega \hertz$ with 12-bits ADC resolution; digitized data are zero suppressed, merged, packed by an onboard FPGA and sent optically to the event builder using a custom protocol. The core of the board is the large FPGA (MicroSemi SmartFusion2 SM2159T), that handles the ADCs protocol and timing, sparsifies and compresses the digitized data and forms a packet that is sent to the following card or to the DAQ servers through optical fibers.

\section{Qualification tests}
The harsh operating environment requires to test and qualify all the components: radiation damage and operation in magnetic field (especially for components like DCDC converters, containing inductive elements). Numerous tests campaigns are needed to qualify a complete set of components.

The adopted Microsemi SmartFusion 2 SM2150T FPGA is already qualified by the producer as SEL free and for a very low SEU probability.

The ADC and DCDC converter has been tested with neutron irradiation (performed at the ENEA Frascati Neutron Generator) and gamma irradiation (performed at the ENEA Calliope facility).

The ADC has been tested up to $20\ \kilo\text{rad}$ and $6 \ \times 10^{12}\ 1\ \text{MeV (Si) n}/\centi\meter^2$ digitizing a $200\ \kilo \hertz$ sinusoidal signal and converting it back to the analog signal. We have analyzed more than $300\ \text{GB}$ of data from both tests and we have observed no evidence of bit flips or waveform shape variation\cite{Qualif}.

The DCDC converter has been tested up to $20\ \kilo\text{rad}$, observing an increase of the output voltage of about $0.5\ \volt$ at $20\ \kilo\text{rad}$. An irradiation of $0.5 \ \times 10^{12}\ 1\ \text{MeV (Si) n}/\centi\meter^2$ causes an increase of $40\ \milli \volt$.

The DCDC converter has been tested also in a magnetic field up to $1.5\ \text{T}$ at the INFN Lasa laboratory. The test shows that the current tends to increase slightly with the increase of the applied magnetic field, with a consequent decrease in conversion efficiency. The device provides still the same output voltage.

\section{Conclusion}

The Mu2e calorimeter working environment sets several constrains in the digitizer design. At the present the first Mu2e waveform digitizer has been defined and designed. All the relevant components have been tested and qualified to operate in high magnetic field and to survive to a ionization dose of $20\ \kilo \text{rad}$ and a neutron flux of $6 \ \times 10^{12}\ 1\ \text{MeV (Si) n}/\centi\meter^2$. 

A new version of the digitizer has been desined and a new qualification test campaign is planned at Helmholtz Zentrum Dresden Rossendorf in June 2018.

\section*{Acknowledgments}
We are grateful for the vital contributions of the Fermilab
staff and the technical staff of the participating institutions.
This work was supported by the US Department of Energy;
the Italian Istituto Nazionale di Fisica Nucleare; the Science
and Technology Facilities Council, UK; the Ministry of Education and Science of the Russian Federation; the US National Science Foundation; the Thousand Talents Plan of China; the Helmholtz Association of Germany; and the EU Horizon 2020 Research and Innovation Program under the Marie Sklodowska-Curie Grant Agreement No.690385 and No.734303. Fermilab is
operated by Fermi Research Alliance, LLC under Contract No.
De-AC02-07CH11359 with the US Department of Energy, Office of Science, Office of High Energy Physics.

\end{document}